\documentclass[12pt]{article}
\usepackage{setspace}
\usepackage[T1]{fontenc}
\usepackage[latin9]{inputenc}
\usepackage[dvips]{color}
\usepackage[english]{babel}
\usepackage[pdftex]{graphics}
\usepackage{graphicx}
\usepackage{amssymb}
\usepackage{amsmath}
\usepackage{amscd}
\usepackage{enumerate}
\usepackage{epsfig}
\usepackage{amsfonts}
\begin{document}
\title{Evolutionary Foundations of Mathematics}
\author{Ruhi Tuncer}
\date{
\begin{flushleft}
\footnotesize{$^{}$  \textit{Department of Economics}, Galatasaray University, . Ciragan Cad. No:36, 34357 Ortakoy, Istanbul, Turkey. Tel:+902122274480-244 Fax:+902122582283. E-mail addresses: tuncer.ruhi@gmail.com 
}
\end{flushleft}
}
\maketitle
\baselinestretch \renewcommand{\baselinestretch}{1.2}
\renewcommand{\abstractname}{Abstract}
\bibliographystyle{plain}
\vspace{-0.4in}
\footnotesize
\begin{abstract} We propose a simple cognitive model where qualitative and quantitative comparisons enable animals to identify objects, associate them with their properties held in memory and make naive inference. Simple notions like equivalence relations, order relations are used. We then show that such processes are at the root of human mathematical reasoning by showing that the elements of totally ordered sets satisfy the Peano axioms. The process through which children learn counting is then formalized. Finally association is modeled as a Markov process leading to a stationary distribution. 
\end{abstract}
\footnotesize
\vspace{-0.in}
\vspace{-0.in}
\newline

\normalsize
\vspace{0.3in}
\begin{doublespace}	

\section {Introduction}

The human race is a product of evolution. All our abilities have come down through generations as a result of natural selection. The basic human ability of "making mathematics" is no exception. So it should be reasonable to search for its antecedents in the animal kingdom.
	
At first sight, searching for mathematical ability in animals might seem to be surprising. After all, animals are not supposed to be intelligent. Intelligence is monopolized by the human species and mathematics is one of its supreme achievements. Animal behavior is supposed to be governed by instinct. However, anyone who has watched a documentary can not fail to make certain observations. Imagine predators and their prey. The predator seems to know its own speed and the speed of its prey. In most cases prey can move faster than predators so that predators approach their prey under cover. Once the predator is in sprinting distance it attacks catching its prey. In such a case a comparison of speeds and distances is required to make a successful attempt. Such quantitative comparisons seem to be at the basis of our own mathematical ability. 
	
Land animals live in an essentially two dimensional space. Sea creatures, tree dwelling animals and most birds live in a three  dimensional space. Living and moving in these spaces necessitate cognitive abilities like speed and distance recognition. Factors like predator prey relations, access to food sources, force animals to make quantitative comparisons. The notions of "more" and "less" , "before" and "after" are essential for survival. All animal species have evolved to the point of possessing cognitive abilities which enable them to survive in geometric space. The human species is no exception. 
	
Our aim in this paper is to present a mathematical model of the cognitive processes that enable animals (including humans) to compare objects and quantities. We emphasize the expression "mathematical model". Any scientific study of animal and human cognitive processes has to be based on empirical research. The study of the neurological structures which underly these cognitive processes is beyond our capability. We will content ourselves with constructing a model of how qualitative and quantitative comparisons are made and structured. We will use the notation of symbolic logic and elementary notions of equivalence relations, truth, order relations by showing how they arise intuitively through qualitative and quantitative comparisons. We will then show how naive inference can be made by association.
 As the title shows we  are going to argue that human mathematical reasoning is a highly developed form of these processes and that work on the foundations of mathematics has to take into account the evolutionary approach. 

There is an extensive literature on the evolutionary and cognitive foundations of mathematics. Good reviews of literature can be found in Beran (2008) , Hauser and Spelke (2004) and Feigenson, Dehaene and Spelke (2004) in the context of cognitive neurosciences. In the field of cognitive architectures related approaches exist. A good review of literature can be found in Langley, Laird and Rogers (2009). However a formal cognitive model of how mathematical ideas come about is lacking.

\section{The Model}

A cognitive system enables animals to become aware of the outside world and adjust their behavior accordingly to survive. Any cognitive system has to have perception, memory and the ability to make inference. Information about objects and their behavioral patterns perceived  by the the cognitive unit are stored in memory in the form of images.  Different units of information i.e. images, are compared and combined to make inference in order to draw valid conclusions about the external world. Comparisons are made by superposition of images. Combining information is made through association. 

The world  consists of objects and distances between them. Objects have certain observable defining physical characteristics.
\bigskip

\textbf{Definition: Objects}
An object is a set of characteristics. We denote objects by  ${x_i} $ and their characteristics by $\{{c_{i1},...,c_{in}\}}$ for $i\in N$.

If the defining characteristics of two objects match, then these objects are said to be similar. We express the notion of similarity in the following way.
\bigskip

\textbf{Definition: Similarity}
 When $c_{i1}\equiv c_{j1} ...c_{im}\equiv c_{jm}$, objects ${x_i} $  and ${x_j} $  are similar for $i\neq j$ and $m\leq n$ where $\equiv$ denotes identity and the numbers are just name tags.  We denote similarity by $x_{i}\sim x_{j}$.
We denote the set of objects similar to $x_{i} $ by $[X_{i}]$. 
Obviously  $ [X_i] $ is an equivalence class representing say, an animal species. Equivalence classes partition the set of objects.

\textbf{Definition:}
$x_{i}\rightarrow y_{j}$ represents moving from object $x_{i} $ to object $ y_{j}$.
We denote the distance between these objects by $\partial (x_{i},y_{j})$.

The set of characteristics serves as a template to identify a particular object as a member of a species. In the memory,  all species have certain properties and behavioral patterns associated with them. These properties are attributed to the elements of the class representing that species as soon as they are identified as belonging to that class. 
These patterns can be represented by predicates like $P(x_{j})$  for $x_{j}\in  [X_{i}]$. 
Relations between different species can be expressed by binary predicates
like $P(x_{i},y_{j})$ where $x_{i}\in \lbrack X_{i}]$ and $y_{j}\in \lbrack
Y_{j}].$ . Notice that similarity $x_{i}\sim x_{j}$ and distance $\partial (x_{i},y_{j})$ already defined are relations between objects $x_i$ and $y_j$ .

\subsection {Naive Logic} 
Survival depends on making valid inference for most of the time. To draw valid conclusions, information has to be processed through association and comparison. This requires  procedures resembling those of logic. We will try to show how such procedures can be obtained intuitively. To start with, we need a notion of truth. Informally, what is observed and remembered is true. 

\textbf{Definition: Truth}  When the characteristics of objects match, we say that 
$x_{i}\sim x_{j}$ is true. In other words the objects are similar.
\bigskip

\textbf{Definition: Negation}
 is the failure of characteristics to match.
\bigskip

\textbf{Definition:} 
 For $x_{j}\sim x_{i}$ , 
 $x_{j}\in  [X_{i}]$ is true. That is to say, the object is identified.  

\bigskip

\textbf{Definition: Truth for Predicates} 
 For $x_{j}\in  [X_{i}]$, $P(x_{j})$ is true. The object has the properties associated with its species. 

\textbf{Definition: Negation for Predicates} 
$\daleth  P(z_i)$i.e. object $z_i$ does not have property $P$ .
\subsubsection {Combining Information to Draw Valid Conclusions}

Drawing the conclusion that when $x_{j}\in \lbrack X_{i}]$ is true, $%
P(x_{j}) $ is true and that it can not be false is valid. Since the same
behavioral pattern can be associated with different species, the conclusion that
when $y_{j}\in \lbrack X_{i}]$ is false $P(y_{j})$ can be true is equally
valid. Yet for another species $z_{j}\in \lbrack X_{i}]$ can be false and \ $%
P(z_{j})$ can be false. This is exactly what is meant when we use the
implication operator $x_{j}\in \lbrack X_{i}]$ $\supset$ $%
P(x_{j}) $  . Other logical operators like $\vee $ and $%
\wedge $ can be introduced and interpreted in the same way.

The possibility that the same behavioral patterns can be expected from members of different species shows that predicates are in fact, equivalence classes of objects sharing a particular property. Once an object is associated with the properties its species has, objects of other species having the same property come to mind. Association works both ways, from object to predicate and from predicate to other objects for which it is true.  

To formalize

\textbf{Notation:}  $A\Longrightarrow B$ denotes association i.e. $A$ brings $B$ to mind. 

 $x_{i}\in \lbrack X_{i}]  \Longrightarrow P(x_i)$

$P(x_i)\Longrightarrow P(y_j) $

$P(y_j)\Longrightarrow y_j \notin [X_i] $ $\Longrightarrow $ $y_j \in [X_i] $ is false but $P(y_j) $ is true.

On the other hand $y_j \notin [X_i] $ $\Longrightarrow $ $z_j \notin [X_i] $ and $P(z_j)$ can be false.

Once these associations are made and stored in memory, they give rise to the implication operator  

 $x_{j}\in \lbrack X_{i}]$ $\supset$ $P(x_{j}) $  .

There are other ways in which association may occur. 

Let $c_{ik}(x_i) $ denote the characteristic named $k$ of the object $x_i$. One might make  the 

association $c_{ik}(x_i) \Longrightarrow c_{jl}(y_j) $.

Yet another way in which association might occur is the following:

 $c_{ik}(x_i) \Longrightarrow P(y_j) $, the characteristic of an object might be the property of another.

Although the operator $\Longrightarrow$ resembles $\supset$ they are not the same. In

$P(x_i)\Longrightarrow P(y_j) $, the truth of $P(y_j)$ is not necessarily based on the truth of $P(x_i)$. 

For example, the following may happen:

$\daleth A \Longrightarrow A $.

Let $\cal C$ be the set of characteristics and let $\cal P$ be the set of predicates. 

\textbf{Definition:} Association is a mapping from $ \cal C \cup \cal P $ to $ \cal C \cup \cal P $.

That $x_i\sim x_j$ is an equivalence relation can be shown now that we have formulated  the implication relation. 
 
 $x_i\sim x_i$ 

If  $x_i\sim x_j$ then  $x_j\sim x_i$ 

If  $x_i\sim x_j$ and  $x_j\sim x_k$  then  $x_i\sim x_k$ 

When we say for $x_{j}\in  [X_{i}]$, $P(x_{j})$ is true, we are using the universal quantifier $\forall$ implicitly. To introduce the existential quantifier $\exists$ we note that all members of a species are not completely alike. Variations within a species gives rise to the existential quantifier. As an example we can think of red parrots and blue parrots. Color in this case is not a defining characteristic.   

\subsection {Motion and Space}

In the same way, we can define similarity for motion between objects.
If $x_{i}\rightarrow y_{j}$ then $y_{j}\rightarrow x_{i}.$ Moreover $%
\partial (x_{i},y_{j})=\partial (y_{j},x_{i})$.
If $x_{i}\rightarrow y_{j}$ and $y_{j}\rightarrow z_{k}$ then $%
x_{i}\rightarrow z_{k}.$
Clearly what we have here is another equivalence relation. We can define
space as the union of the equivalence relations constituted by motion from
one object to another.

\textbf{Definition: Space}
Space is the union of the equivalence classes of the arrows which
represent the possibility of motion.

We are immediately led to the recognition that order relations exist in
equivalence classes. Imagine the following situation: object $x_{i}$ moves
towards object $x_{k}$. It can pass through \ $x_{j}$ where $j$ may take any value. Moving from $ x_{i}$ to $x_{k}$ involves the choice of a possible path.  Such a choice will be dictated by the comparison of the distances involved.

$x_{i}\rightarrow x_{j_{1}}\rightarrow x_{k}$ is equivalent to $%
x_{i}\rightarrow x_{j_{2}}\rightarrow x_{k}$

However the distances involved are not necessarily the same. This is where a
quantitative comparison is made and the quantitative differences between the
elements of an equivalence class play their part.
\section {Order}
While qualitative comparisons give us equivalence classes partitioning the set of objects, quantitative comparisons give us order relations. 

Order relations are defined as relations having the following properties:

$x_i \preceq x_i $

If $x_i \preceq x_j $ and $x_j \preceq x_k $ then $x_i \preceq x_k$

If $x_i \preceq x_j$ and $x_j \preceq x_i $ then $x_i \approx x_j $ 

where $\approx $ stands for equivalence. Strict ordering is denoted by $x_i \prec x_i $.

\textbf{Assumption: Truth for Ordering}  Cognitive units can compare sizes, distances and realize relations like $x_{i} \preceq x_{j}$ ,  $\partial (x_{i},y_{j})\preceq\partial (x_{i},z_{j}) $ , $\partial (x_{i},y_{j})$ $\approx$ $\partial (x_{i},z_{j})$ as true.

\textbf{Definition: Ordering Time} The notions of before and after are important for cognitive units. We will order time as follows:

$t\preceq \tau $ which obviously means that $\tau$ is later than $t$. 

Since distances can change in time, we will index them. 

 $\partial_\tau  (x_{i},y_{j}) $ $\preceq  \partial_t  (x_{i},y_{j})$ means that for $t\preceq \tau $ 

the distance between $x_i $ and $y_j$ has decreased.

\subsection {Well ordering}

A property of order relations that we are interested in is that of well ordering. A set of objects $X$ is well ordered if all non-empty subsets of $X$ have a first element.
That is to say, $\forall X'\subseteq X$ there exists  $x_i$ such that  $\forall x_{j}\in X' \subseteq X$ the following holds: 

 $x_i$ $\preceq$ $x_j \in X'$.

If a set is well ordered then it is totally ordered i.e. all the elements of the set are comparable. In other words 

$\forall x_{i} , x_{j}\in X $

either $ x_{i} \preceq x_{j}$ or $x_{i}\succeq x_{j}$ or $x_{i}\approx x_{j}$ .

If all the elements of $X$ are not comparable i.e. if there exist elements $x_i$ and $x_j$ such that none of the relations above hold then we have a partial order.

\subsection {Ordering Predicates}

Order relations are defined in so general a way that the implication operator can impose an order on predicates. 

A$\supset$A

If A$\supset$B and B$\supset$C then A$\supset$C.

If A$\supset$B and B$\supset$A then $A$ and  $B$ are equivalent

which gives us a partial order since not all propositions imply one another.

Notice that the relation set-subset is an order relation. 

If $S'\subset S$ we can write $S' \prec S$ meaning that $ S $ contains more elements than $S'$.

Each cognitive unit can be said to have its own ordering of predicates which can be interpreted as its world view.   

The association operator $\Longrightarrow $ that we have defined does not impose an order since 

if $A \Longrightarrow B$ and $B \Longrightarrow A$ 

$A$ and $B$ are not necessarily equivalent. The association relation can best be modeled  as a communication relation in Markov chains.

\section {Typical Scenario}
To show how the setup we have introduced can describe animal behavior we propose the following scenario. 

Let $ \lbrack X_{i}]$ represent a species of predators and $ \lbrack Y_{j}]$ that of prey. Object $x_i$ is perceived by object $y_j$ . Object $y_j$ identifies  $x_{i}$ as a member of the predator species  $ \lbrack X_{i}]$ . So $P(x_i)$ is true. If the predator is moving towards the prey  ,   $x_i$ will attack $y_j$ . In symbols 

$x_{i}\in \lbrack X_{i}] $ $ \supset$  $ P(x_i)$

$\partial_\tau  (x_{i},y_{j}) $ $\preceq  \partial_t  (x_{i},y_{j})$ for $t \preceq \tau $

$P(x_i)$ $\wedge$  $[\partial_\tau  (x_{i},y_{j}) $ $\preceq  \partial_t  (x_{i},y_{j})] $$\supset$  $P(x_i,y_j)$ 

$P(x_i ,y_j)$ $\supset$ $R(y_j)$ where$P(x_i,y_j)$ and $R(y_j)$ stand for $x_i$ attacking $y_j$ and $y_j$ running away.

So
 $x_{i}\in \lbrack X_{i}]$ $\wedge$  $[\partial_\tau  (x_{i},y_{j}) $$\preceq  \partial_t  (x_{i},y_{j}) ]$ $\supset$  $R(y_j)$ .

It is interesting to note that the train of thought is ordered by the implication operator. Qualitative and quantitative comparisons are combined to give a reaction.

\section {Numbers}

In the introduction, we have claimed that quantitative comparisons seem to be at the basis of our own mathematical ability. We have outlined a cognitive model of how quantitative and qualitative comparisons are made and structured. To validate our claim we have to show how the concept of number can arise naturally in the model that we have proposed. In our setup we have modeled quantitative comparisons by order relations. In mathematics the accepted way of defining natural numbers is using the Peano axioms. So now our intention is to show that members of ordered sets satisfy these axioms.  Since the Peano axioms provide a definition of number which is sufficient to construct most of applied mathematics, obtaining them from our setup will validate our claim.  

\subsection*{Peano Axioms}

Every axiomatic system has to start with undefined concepts. In Peano's axiomatic system, one, number and successor are the undefined concepts. The axioms are a list of properties that these undefined concepts have.

One is denoted by {1} , a number by $N$, and a successor by  $S(N)$.

The set of axioms is as follows:

i) 1 is a number.

ii) Every number has a unique successor.

iii) There is no number whose successor is 1.

iv) Distinct numbers have distinct successors.

v) Let $P$ be a property. If $P(1)$ is true and if $P(N)$ $\supset $ $P(S(N))$ 

then $P$ is true for all $N$.

The last axiom is the well-known induction property which is why we have shown how the implication operator arises naturally through association. 

In this system, numbers are abstract mathematical objects satisfying these axioms. The  familiar sequence $1,2,3,...$ is a model for the axiomatic system.  

The operation of addition is not part of the axioms. It is defined on numbers already defined.

\textbf {Definition: Addition} For numbers $N$ and $M$

$N+1=S(N)$

$N+S(M)=S(N+M)$

Although the word successor immediately suggests order, in Peano's system numbers are ordered after they and the operation of addition are defined . In this system number precedes order. However in the approach we propose, order precedes number. 

 \section {Order Relations and Peano Axioms}

To start with we have to define a successor function for order relations.

Let $X$ be a totally ordered set of distinct objects. 

\textbf {Definition: Successor Function}

 $S(x_i)\approx x_j$ is the successor of $x_i$  if for $x_i , x_j , x_k \in X $

$x_j \succ x_i $ and $(\forall x_k \succeq x_i) $  $x_j \preceq x_k$.

The following properties of the successor function will lead us to the result that we want to establish.

\textbf{Lemma 1}  The successor of $x_i$ is unique.

\textbf{Proof}  

Let $x_j$ and $x_l $ be two successors of $x_i$ .

$S(x_i)\approx x_j$  and  $S(x_i)\approx x_l$ .

$x_j \succ x_i $ and $(\forall x_k \succeq x_i) $  $x_j \preceq x_k$ and 

$x_l \succ x_i $ and $(\forall x_k \succeq x_i) $  $x_l\preceq x_k$.

$x_j \succ x_i$ so that $x_l \preceq  x_j $

On the other hand

$x_l \succ x_i $ so that  $x_j \preceq  x_l $ proving that $x_j \approx x_l $.

\textbf{Q.E.D.}

\textbf{Lemma 2}

Distinct objects have distinct successors.

\textbf{Proof}

Suppose the contrary.

$S(x_i) \approx x_j $ and $S(x_m) \approx x_j $.

$x_j \succ x_i$ and $x_j \succ x_m $.

Now $ (\forall x_k \succeq x_i) $ $x_j \preceq x_k $ and

$ (\forall x_k \succeq x_m) $ $x_j \preceq x_k $ .

We have a totally ordered set so that either $x_m \succeq x_i$ or $x_m \preceq x_i$.

If  $x_m \succeq x_j $ since  $x_j \preceq x_m $ equivalence is proved.

If $x_m \preceq x_j $ then $x_j \preceq x_i$ which is impossible since $x_i \preceq x_j$.  

\textbf{Q.E.D.}

We can now prove a theorem.

\textbf{Theorem}

The elements of a well ordered set of objects satisfy Peano's axioms.

\textbf{Proof}

Every object in the set will be called a number.

Distinct objects will have distinct numbers.

The first element of the well ordered set will be called the number 1.

Lemma 1 shows that the successor of each object is unique.

Lemma 2 shows that distinct objects have distinct successors. 

There is no object whose successor is the first object since the first object is the minimal element of the well ordered  set.

The induction property is satisfied since well ordered sets satisfy the transfinite induction property.

\textbf{Q.E.D.}

Essentially, numbers are names. What counts is the relation between these names. We will express the relation as follows:

Each object is assigned a name N. Distinct objects have distinct names. The names of objects will be ordered in the same way as the objects they represent. The successor of N is the name of the object that is the successor of the object that N represents. 

The relation between names is called addition and is defined as follows:

For names N and M

$S(N)=N+1$

$N+S(M)=S(N+M)$ .

This much is enough to give us counting and adding up the number of elements in subsets of the ordered set. 

\section {Children and Counting }

There is a huge literature on how small children learn to count. In developed cultures names for numbers exist and children start counting counting at around  age two. However despite the fact that they count in the correct order, they do not know the exact values of the numbers that they have memorized.  

"Although she uses the counting words correctly in the count routine, she evidently interprets each word above one as simply meaning "more than one." With months of counting experience, as well as other cognitive advances that are running in parallel, children progress from understanding the meaning of "one" to understanding "two," and then "three"; this progression is highly systematic with no evidence of children learning other numbers in the integer count list first, nor learning the meaning of three before they learn the meaning of two (Wynn, 1990). After this slow, systematic, stepwise progression, children take a leap forward.They form the induction that each word in the counting routine gives the cardinal value of a set composed of a specific number of individuals, that each word denotes a set with one more individual than the previous word, and that the succession of cardinal values picked out by the number words can be continued indefinitely, with no upper bound."  (Hauser and Spelke 2004).

It seems reasonable to formalize this process in our setup. Children memorize the set of numbers in the correct order.

$\{one, two, three, four, five, ...\}$. Then they are faced with a set of objects $\{a, b, c,d, e,...\}$ where the objects are completely alike. 

First they learn to associate the number one with the set $\{a\}$.

$\{ one\} \Longrightarrow \{a\} $ a set having a single element.

Faced with $\{ two\}$ they interpret the number as more than one: $\{ two\} \Longrightarrow \{a, b, c,d, e,...\}$ i.e. $two \succ one$ since $ \{a\} $ $ \prec $ $\{a, b, c,d, e,...\}$.

 Then they progress to understanding the meaning of  $\{ two\}$.

 $\{ two\}\Longrightarrow \{a, b\}$.

Faced with $ \{ three\}$ they interpret it as more than $\{two \}$  i.e.

 $ \{ three\} \Longrightarrow  \{a, b, c,d, e,...\}$ and $ three \succ two $

  since $\{a, b \} \prec \{a, b, c,d, e,...\}$.

When they realize that $ \{ three \} \Longrightarrow \{a,b,c \}$ they are ready to take the crucial step.

$ \{a\} \prec \{a,b\} \prec \{a,b,c \} $ so

$\{ one \prec two \prec three \}$ $\Longrightarrow$  $\{ \{a\} \prec \{a,b\} \prec \{a,b,c \}\}$

taking the inductive step they have

 $\{ one \prec two \prec three \prec ...\prec N \}$ $\Longrightarrow$  $\{ \{a\} \prec \{a,b\} \prec \{a,b,c \}\prec ...\prec \{a,b,c,...,N \}\}$.

Using the successor function defined on subsets

$ S(\{a\}) \approx \{a,b\} $

 $ S(\{a,b\}) \approx \{a,b,c\} $

$ S(\{a,b,c,...,M\}) \approx \{a,b,c,...,N\} $ .

Now 

$\{ one\} \Longrightarrow \{a\} $ the word one is associated with all single objects.

$S(one)\approx \{two\} \Longrightarrow S(\{a\}) \approx  \{a, b\}\approx \{a\} \cup   \{b\}  $ and 

$S(two) \approx  \{three\} \Longrightarrow S( \{a, b\}) \approx \{a,b,c\} \approx \{a,b\} \cup   \{c\}  $.

$S(M) \approx  \{N\} \Longrightarrow S( \{a, b,c,...,M\}) \approx \{a,b,c,...,M,N\} \approx \{a,b,c,...,M\} \cup   \{N\}  $.
 
In this context union $\cup$ should  be interpreted as  including "more and more" objects

 in  a given collection.  The objects are completely alike. The word $\{ one\}$ is associated 

with all single objects. 

Each set in the ordered sequence

 $\{ \{a\} \prec \{a,b\} \prec \{a,b,c \}\prec ...\prec \{a,b,c,...,N \}\}$

includes "one" more  object than the preceding set and the sets are

 increasing one by one so that the numbers associated with them  have to increase

one by one.

Including one more object to a collection gives us addition i.e.

$S(M)  \Longrightarrow \{a,b,c,...,M\} \cup   \{N\}  $ and

$S(M) \approx N \approx M+1$ .

$ \{a,b,c,...,M\} \cup \{a,b,c,...,N\} \Longrightarrow M+N $ hence

$ S(M) +N \approx S(M+N) $.

This way of generating natural numbers is similar to Von Neumann's set-theoretic definition of numbers. However the crucial difference is that in this approach order precedes number whereas in Von Neumann's approach number precedes order.

\section {Association as a Markov Chain}

The preceding arguments have a deterministic flavor. In reality the processes in question are stochastic. The association operator  $\Longrightarrow $  can be modeled as a Markov process. The stationary distribution of such a process would lead to a possible ordering of predicates, giving a cognitive picture of the world for each cognitive unit. 

To formalize let $P_i $ and $P_j $ be two predicates. Each predicate that comes to mind can be considered as a state in a state space. 

Let $\{V_n , n=0,1,2,...\} $ be a stochastic process. The set of possible values for this process is the set of nonnegative integers. If $ V_n =i$ , the process is said to be in state $i$ at  time $n$ (Ross 1996). We will say that when $ V_n =i$ the cognitive unit has  in mind the predicate $P_i$ .

We assume that $V_n$ is a Markov process

 i.e. $p(V_{n+1}=j |V_n =i, V_{n-1} i_{n-1} ,...,V_0=i_0) =p((V_{n+1}=j |V_n =i)$.

Let $p_{ij} $ denote the transition probability $p((V_{n+1}=j |V_n =i)$  . At time n the cognitive unit has $P_i $ in mind which it associates with  $P_j $ i.e.   

$p((V_{n+1}=j |V_n =i)=p(P_j|P_i) $ .

Let $p_{ij}^n =p(V_{n+m} =j |V_m =i)$ i.e. the n step transition probability of passing from state $i$ to state $j$. These probabilities can be computed using the Chapman-Kolmogorov equations
$p_{ij}^{n+m} = \sum_k p_{ik}^n p_{kj}^m $  . In our context, these probabilities will represent passing from  $P_i $ to $ P_j $ in n steps.

State $j$ is said to be accessible from state $i$ if $ p_{ij}^n > 0$ . If both  $ p_{ij}^n > 0$ and  $ p_{ji}^m > 0$, states $i$ and $j$ are said to communicate. In other words $P_i \Longrightarrow P_j $ and $P_j \Longrightarrow P_i $. Communication is an equivalence relation so that we have equivalence classes of predicates which are associated with one another. In this way we can impose a temporal partial order on predicates using the association operator.

If starting at state $i$ the process returns to that same state with probability one, $i$ is called a recurrent state. States which communicate with $i$ are also recurrent so that recurrence is a class property.  A Markov chain is irreducible if all states communicate with one another. 

The following result will be useful for our purposes: for an irreducible aperiodic Markov chain if $\pi_j = lim_{n \rightarrow \infty }$   $ p_{ij}^n > 0 $ then  all states are positive recurrent and $\{ \pi_j , j=1,2,...\} $ is the unique stationary distribution of the process i.e. the solution of the equation

$\pi _j = \sum_k \pi_k p_{kj}$ .

 Putting $p_j = \frac{\pi_j}{\sum_k \pi_k }$ we have the unconditional probabilities that the system is in state $j$ i.e. $p_j = p(V=j) =p(P_j)$ (Ross 1996). Since this is the stationary distribution, these unconditional  probabilities are independent of time. They  give us the proportion of time that the cognitive unit has in mind the predicate $P_j$.

What we have here is an instrument that permits us to classify predicates. States that communicate i.e. predicates that are associated with one another form equivalence classes. With the passing of time associations settle at their steady state  $\{p(P_j) , j = 1, 2, ...\} $ and we have a unique pattern of having in mind information expressed as predicates. 
Each cognitive unit has its own pattern which largely characterizes the functioning of the cognitive system.

Cognitive units can and do confuse association, implication and causality. In principle $ P_i \Longrightarrow P_j $ has nothing to do with the truth of the predicates $P_i $ and $P_j $ .
Whether the implication $ P_i \supset P_j $  following   the association $ P_i \Longrightarrow P_j $ will be valid i.e. true depends on their truth values. Once valid implications are made, they will impose an order on  predicates which will characterize their world view.  

\section {Indications for a Prolog program}
The following indications can be helpful to simulate the model we have proposed.

In the model, the basic concepts are equivalence relations and order relations. The well known Animal Identification game in Prolog is ideal for simulating what we have called similarity and to form equivalence classes. The association    $x_{i}\in \lbrack X_{i}] $ $ \Longrightarrow P(x_i)$ can be made using the same game. Then $ y_j \notin \lbrack X_{i}] $ can be searched for which  $P(y_j )$ is true. Finally finding $z_j \notin [X_i] $ for which $P(z_j) $ is false will give us the implication rule $ x_i \in [X_i ] \supset P(x_i ) $.

To express order relations we can use the well known genealogy programs in every Prolog tutorial. Replacing the terms for family members with "more" and "less" would do the job. 

The process through which children learn counting can be simulated in the following way. We can define the ordered sequences 

$ \{ \{a\} \prec \{a,b\} \prec \{a,b,c\} \prec ...\prec \{a,b,c,...,N\} \}$ and 

$ \{one , two, three, ... \}$.

Make the first association $ \{one\} \Longrightarrow \{a\} $. 

Then $ \{two\}$ is associated with "more" than $ \{one\}$ i.e. $ \{one\} \prec  \{two\}$.

However $ \{two\} \Longrightarrow \{a,b.c,d,...\} $

Checking the truth of this statement the program will give the result  "false". Then until the association  $ \{two\} \Longrightarrow \{a,b\} $ is made the result will be declared "false". The same process will be repeated until the correct isomorphism between objects and number words is obtained.

Then the successor function can be defined inductively using the union operator of Prolog.

$S(\{a\}) \approx \{a,b\} \approx \{a \} \cup \{b\} $ 

$S(\{a,b,c,...,M\} \approx \{a,b,c,...,M\}\cup \{N\} \approx \{a,b,c,...,N\}  $.
 
\section {Conclusion}

We have outlined a simple cognitive model of how qualitative and quantitative comparisons are made and structured. Qualitative comparisons give rise to naive logic through association. Quantitative comparisons lead to order relations. We have shown the possibility of constructing natural numbers by proving that totally ordered sets satisfy the Peano axioms. We  have then shown how children learn counting through association. Association is then modeled as a Markov process leading to a stationary distribution which characterizes the mindset.

\end{doublespace}
\end{document}